\documentclass[12pt,preprint]{aastex}
\begin{document}
\title{The Difference Between the Amati and Ghirlanda Relations}
\author{Amir Levinson\altaffilmark{1} \& David Eichler\altaffilmark{2}}
\altaffiltext{1}{School of Physics and Astronomy, Tel-Aviv University, 
Tel Aviv 69978, Israel; Levinson@wise.tau.ac.il}
\altaffiltext{2}{Physics Department, Ben-Gurion University,
Beer-Sheva 84105, Israel; eichler@bgumail.bgu.ac.il}
\begin{abstract}
It is pointed out that the beaming correction commonly inferred from 
the achromatic breaks in the afterglow light curve, is biased in situations where
the isotropic equivalent energy is affected by factors other than 
the spread in opening angles.   In particular, it underestimates the beaming
factor of sources observed off-axis.   Here we show that both the slopes and scatters
in the Amati and Ghirlanda relations, and the difference between them, are quantitatively
consistent with a model proposed recently, in which the $E_{iso}-\nu_{peak}$
relation, as originally derived by Amati et al., is due to viewing angle effects.
The quantitative difference between them confirms the relations between opening angle and
break time suggested by Frail et al.  

\end{abstract}


A correlation between the isotropic equivalent energy of GRB's, $E_{iso}$, and the
location of the spectral peak at local redshift, $\nu_p$, has been reported 
recently for a sample of BeppoSAX and HETE2 sources 
with measured redshifts and well observed spectra (Amati et al.2002; Atteia et al. 2003; 
Lamb et al. 2004).  This relation can be represented as a power law,
\begin{equation}
(E_{iso}/10^{52} \rm{erg})\sim (h\nu_{p}/100 KeV)^{\alpha},
\label{Enu}
\end{equation}
with $\alpha=2$, and spans two orders of magnitude in $\nu_{p}$,
roughly from 10 KeV to 1  MeV,  and about four orders of magnitude
in $E_{iso}$. The extension to low energies is based on a small
number of X-ray flashes with measurable redshifts. It does not
preclude the possibility that many X-ray flashes derive their low
apparent peak luminosities in part due to cosmological redshift.
A similar correlation between the collimation-corrected energy and $\nu_p$, 
with a smaller scatter and a different slope, has been reported by 
Ghirlanda et al. (2004; hereafter GGL04) for a subsample of the sources that exhibit 
achromatic breaks in the afterglow light curves.  

Eichler \& Levinson (2004; hereafter EL04) proposed that the $E_{iso}-\nu_p$ correlation can be 
interpreted as due to an orientation effect.   They have shown that an annular jet with a
single, universal spectrum can give rise to the observed relation over about two
orders of magnitude in $\nu_{p}$, when observed off-axis along different sight lines, 
provided the opening angle and angular width 
of the jet are larger than $1/\Gamma$, where $\Gamma$ is the Lorentz factor of the 
emitting material (see appendix for an approximate analytic derivation).   The effect is not 
necessarily specific to annular jet; 
it applies to any geometry in which the emission along the line of sight is comprised 
of contribution from an extended source (e.g., Toma, et al. 2005). However, there are {\it a priori} reasons to favor 
an annular geometry (see below).
EL04 also calculated the rate distribution of observed peak energies
$\nu_{p}$, and have shown that the relative number of XRFs observed is compatible 
with the model, provided the angular width of the emitting region is not too large. 
(Off-axis effects have been considered also by Yamazaki et al. [2004, and 
references therein].  However, they 
assumed an {\it ad hoc} intrinsic relation between $E_{iso}$ and $\nu_p$ designed to 
fit the observations.)   If off-axis effects are affecting $\nu_p$ and $E_{iso}$ then 
the various quantities one derive from GRB observations need to have this effect folded in.
For example, Eichler \& Jontof-Hutter (2005) have recalculated the average prompt gamma 
ray to blast wave energy ratio and found it to be considerably higher than previous 
estimates, simply because the prompt gamma rays are more affected by viewing angle effects 
than the blast wave energy as inferred from the afterglow.
The normalization of the $E_{iso}-\nu_p$ curve produced by the model of EL04 depends on the
amount of energy per solid angle ejected by the source, which for a standard energy
release is inversely proportional to the solid angle subtended by the emitting region.  Evidently,
any spread in this parameter in a sample of sources should be reflected as 
a scatter in the resultant correlation.  Under the assumption of a standard energy output, 
this scatter can be reduced essentially by correcting for beaming.  If an absolute 
way can be found to correct for beaming, then the model naively predicts the collimation-corrected 
energy to satisfy the same relation as the isotropic equivalent energy, but with a different 
normalization and virtually no scatter.

An attempt to derive a relation between the collimation-corrected GRB energy, denoted here
by $E_{\gamma,app}$ for future purposes, and the 
peak energy has been made recently by GGL04.  These authors
considered a sample of 24 GRBs with measured redshift and peak energy,
for which the break time of the afterglow light curve, $t_{break}$, is well constrained.   
They found indeed a tight relation between $E_{\gamma,app}$ and the observed
peak energy $\nu_p$, but with a different slope: $E_{\gamma,app}\propto \nu_p^{0.7}$. 
At first sight this seems to be in conflict with the prediction of the model discussed above.
However, careful examination reveals that this result is in fact consistent with the hypothesis 
that the Amati relation
is due to viewing angle effects, and that the change in slope is primarily due to a bias
in the beaming correction used by GGL04.

In order to correct for beaming GGL04 estimate the opening angle of the 
emitting jet for each source in their sample, using the two observables, $E_{iso}$ and 
$t_{break}$, and the relation:
\begin{equation}
\theta(t_{break},E_{iso})=0.161\left({t_{break,d}\over 1+z}\right)^{3/8}
\left({n\eta_{\gamma}\over E_{iso,52}}\right)^{1/8}.
\label{thet}
\end{equation}
Here $t_{break,d}$ is the break time measured in days, $\eta_{\gamma}$ is the radiative 
efficiency, and $n$ the density of the circumburst ambient medium, assumed to be constant.
Since the radiative efficiency is unknown, they invoked a single value for all sources of $\eta_{\gamma}=0.2$.
For the circumburst density they adopted the value $n=3$ cm$^{-3}$ for all sources, except five for which
estimates for $n$ are available in the literature.
The collimation-corrected energy was then taken to be: $E_{\gamma,app}=(1-\cos\theta)E_{iso}\simeq (\theta^2/2)E_{iso}$.
Since the opening angle given by eq. (\ref{thet}) depends on $E_{iso}$, it is clear that it may be subject
to biases in cases where $E_{iso}$ is influenced by effects other than the spread in
the opening angle.  This is particularly true for sources observed off-axis, for which
eq. (\ref{thet}) underestimates the beaming factor.   Furthermore, the break time of the 
afterglow emission may appear longer for off-axis observers (Granot et al. 2002), leading to an even 
larger bias.  The latter effect is typically small, except for the 
very soft sources (i.e., those observed at the largest viewing angles).    
In the following we provide a quantitative treatment of these biases.

The model outlined in EL04 assumes a uniform, axisymmetric jet
of opening angle $\theta_2$, with intrinsic spectral peak at
$\nu^*$, and with a hole of angular size $\theta_1$ cut out of it.   This
symmetry was chosen strictly for convenience, and it can be easily seen that
the scaling derived in EL04 applies to more complicated geometries.
The peak frequency and the observed isotropic energy were calculated numerically 
for different viewing angles inside the hole and outside the jet, and it was found that 
a relation $E_{iso}\propto \nu_p^{\alpha}$ holds over a wide range of $\nu_p$, with 
$\alpha$ laying in the range between 2 and 3, depending on source parameters.  In particular,
$\alpha\simeq 2$ when $\theta_2$ and  $\Delta\theta=\theta_2-\theta_1$ are larger than $\Gamma^{-1}$.
The analysis of EL04 does not account for a possible spread in source parameters.  In particular, the 
normalization of the $E_{iso}-\nu_p$ relation produced by the model, depends on the angular distribution of 
the released energy.  To illustrate how this might affect the observed $E_{iso}-\nu_p$ relation,
we consider a sample of sources having a universal spectrum and a standard energy output, $E_{\gamma}$, but 
a range of opening angles.  Denote
by  $A_b=4\pi/\Delta \Omega$ the corresponding beaming factor,
where $\Delta \Omega$ is the solid angle subtends by the gamma-ray emission region.
For a symmetric (double sided), hollow jet we have $\Delta\Omega=4\pi\int_{\theta_1}^{\theta_2}\sin\theta d\theta
=4\pi(\cos\theta_1-\cos\theta_2)$, and 
$A_b=(\cos\theta_1-\cos\theta_2)^{-1} \simeq 2/(\theta_2^2-\theta_1^2)$. 
The isotropic equivalent energy measured for a source observed at some angle 
corresponding to an observed peak energy $\nu_p$,
can then be expressed as,
\begin{equation}
E_{iso}=A_bE_{\gamma}(\nu_p/\nu_{p,max})^{\alpha}.
\label{Eiso}
\end{equation}
Here $\nu_{p,max}=2\Gamma\nu^*$ denotes the peak frequency of a source observed head-on.
Consequently, a sample of sources with fixed $E_{\gamma}$ and a range of beaming
factors $A_b$ would form a family of parallel lines in the $E_{iso}-\nu_p$ plane,
with each line corresponding to a subset of sources with beaming factors
in the interval $(A_b, A_b+dA_b)$.  It is tempting to interpret figure 2 in Ghirlanda et al. (2005)
and figure 5 in Bosnjak et al. (2005) as such.  The sample studied originally by Amati (2002),
that contained particularly bright bursts, may consist mainly of the subset of sources with the 
largest beaming factors, and therefore defines a boundary in the $E_{iso}-\nu_p$ plane,
as suggested by Nakar \& Piran (2004).

There are several reasons why the GRB fireball might be shaped like a thick annulus. 
First, the collimation of supersonic material can itself produce
an annular shaped, high entropy jet (Eichler 1982, Levinson \& Eichler 2000; 
Begelman \& Blandford, private communication) because the jet material tends to accumulate into a
shock-compressed layer at the confining walls. Second, the soft gamma ray emission 
is likely to originate from that part of the jet that is baryon loaded, and the baryon
loading may come from the periphery.  In the particular case of baryon
loading by neutron leakage from the walls of a confining wind or
stellar envelope (Levinson \& Eichler 2003), the neutrons are
quickly charged by collisions near the walls, before they can
penetrate to the center. It was shown in the above reference that
the annular region that is significantly loaded can have a solid
angle that is not too much less than the inner hollow region.
Third, gamma-rays emitted by a compact photosphere (Eichler \& Levinson, 2000) can mostly
impinge on the baryon rich periphery of the outflow and be dragged along by it, and
thus be concentrated into a an annulus with a smaller total solid angle than that
into which they were originally emitted.
The inner core may be comprised of Poynting flux that contribute very little to the 
soft gamma-ray emission, but nonetheless carry a considerable fraction of the ejected 
energy.  Thus, even though the gamma-ray emission region may be annular, the total energy carried
by the outflow is likely distributed more uniformly inside the cone.
Now, the structure of the blast wave driven by the collision of the 
fireball and the surrounding gas depends mainly on the angular distribution of total
energy of the piston.  It is therefore reasonable to assume that at small viewing angles 
the break in the afterglow light curve is associated with the opening 
angle of the outer cone.  At large viewing angles, the break time of the 
afterglow emission may be altered.   The exact shape of the afterglow 
light curve as viewed by off-axis observers may depend on details.  For sight lines 
outside the jet, one might naively expect a break when the Lorentz factor drops to 
$\Gamma\sim 1/\theta_n$, where $\theta_n$ is
the viewing angle measured with respect to the jet axis (Granot et al., 2002).  The angular
separation corresponding to an observed peak energy $\nu_p$ is given by:
$\theta_n-\theta_2=\Gamma^{-1}(2\nu_{p,max}/\nu_p-2)^{1/2}$.  The distribution of opening 
angles inferred from the achromatic breaks of the afterglow emission peaks at 
$\theta \sim 0.1$ (Frail et al. 2001; Guetta et al. 2005; Ghirlanda et al. 2005).  Adopting this value for 
$\theta_2$ we find that $\theta_n-\theta_2<\theta_2$ if $\nu_{p,max}/\nu_p<\Gamma^2/200$.
For reasonable values of the Lorentz factor $\Gamma$ this condition is satisfied essentially
for all sources in the GGL04 sample, and we therefore anticipate the break time not to be
significantly altered by viewing angle effects.  For observers looking down the hole the 
break time may appear unaltered.
Let us denote by $q(\nu_p)=t_{break}(\nu_{p})/t_{break}(\nu_{p,max})\ge1$ the ratio of 
break times measured by off-axis and head-on observers.  Suppose that 
for a source observed head-on, the opening angle obtained by substituting
the observed energy, $E_{iso}=A_bE_{\gamma}$, and break time, $t_{break}$, into eq. (\ref{thet}) 
is roughly $\theta_2$, viz., 
$\theta(t_{break}(\nu_{p,max}),A_bE_{\gamma})\simeq\theta_2$.  Then for the same source observed off axis 
the inferred opening angle would be, $\theta(t_{break},E_{iso})
\simeq \theta_2q^{3/8}(\nu_p/\nu_{p,max})^{-\alpha/8}$, where eqs (\ref{thet}) and (\ref{Eiso}) 
have been used.
The apparent collimation-corrected GRB energy, $E_{\gamma,app}$, is then given by 
\begin{equation}
E_{\gamma,app}={\theta^2\over 2} A_bE_{\gamma}(\nu_p/\nu_{p,max})^{\alpha}
\simeq{q^{3/4}\over1-(\theta_1/\theta_2)^2}E_{\gamma}(\nu_p/\nu_{p,max})^{3\alpha/4}.
\label{Eapp}
\end{equation}
For $\alpha=2$ and $q=1$, the last equation yields:
\begin{equation}
E_{\gamma,app}\propto \nu_p^{1.5},
\end{equation}
in agreement with the result obtained by GGL04, $E_{\gamma,app}\propto \nu_p^{1.416\pm0.09}$.
We expect the index to be somewhat smaller than $1.5$, owing to break time effects contributed by the 
softest sources, those 
with $\nu_p<<\nu_{p,max}$, for which the ratio $q$ may already be large enough to affect
the correlation.   As seen from eq. (\ref{Eapp}) the value of $E_{\gamma,app}$ predicted by 
the model depends in addition on the ratio $\theta_1/\theta_2$.  This parameter
is uncertain in the present picture.  A spread in this parameter
will contribute a scatter in the observed $E_{\gamma,app}-\nu_p$ relation.  The inverse
correlation between $E_{iso}$ and jet opening angle (Frail et al.
2001; van Putten \& Regimbau 2003), seems to imply that 
the annulus is reasonably thick, such that it subtends a
significant fraction of the solid angle subtended by its outer
cone.  In this case the scatter is expected to be rather small.  For the range of 
parameters adopted in EL04 the values of the coefficient $1/[1-(\theta_1/\theta_2)^2]$ 
in eq. (\ref{Eapp}) vary between  1 and 1.5. 

The above theoretical explanation is consistent with the observed correlation between $\nu_p$ and 
the inferred opening angle (Eichler \& Jontof-Hutter, 2005), where the inferred opening angle shows the 
tendency to increase with decreasing $\nu_p$.  The point is that the inferred opening angle  is weakly biased
towards larger values when $E_{iso}$ is underestimated due to off angle viewing (see eq. \ref{thet}).

Whether the observed relations discussed above hold for the entire population of GRBs or are
the result of some selection effects is at present under debate.  Several groups
have attempted to perform consistency checks for large samples of BATSE sources
with known fluence and well determined peak energy.   Nakar \& Piran (2004) analyzed
trajectories in the $E_{iso}-\nu_p$ plane obtained for sources with a measured fluence 
and observed peak energy, $\nu_{p,obs}=\nu_p/(1+z)$, by varying the redshift.  They then define 
a source to be an outlier if the minimum distance between its trajectory and the curve representing 
the Amati relation exceeds a certain value.  By applying this test to a sample 
of bright bursts from (Band et al. 1993; Jimenez et al. 2000) they concluded that 
about 50\% of the sources in their sample are outliers.  However, they also concluded that the line
representing the Amati relation defines a boundary of the region 
in the $E_{iso}-\nu_p$ plane which is populated with sources. In other words, all outliers should be dim, hard 
bursts. As argued above, this boundary may represent the subsample of GRBs with the smallest opening angles.
Band \& Preece (2005) extended this work to a larger sample of BATSE sources and
tested in addition the Ghirlanda correlation.   They find an even larger number of sources 
to be inconsistent with the Amati relation.  The number of outliers to the Ghirlanda relation
depends on the assumed distribution of beaming factors, and appears to be much smaller, at
least for certain choices of beaming correction.   This better agreement may be due to the smaller
scatter in the Ghirlanda correlation, although we are aware of the large uncertainty in the distribution 
of opening angles that might affect the significance of this result.  Different consistency checks 
have been performed subsequently by two other groups.  Ghirlanda et al. (2005) analyzed a sample of 442
bursts with what they term ``pseudo redshifts'' which have been estimated using the 
lag-luminosity relation (Norris et al. 2000; Norris 2002).  Bosnjak et al. (2005) tested the 
consistency of the Amati relation with the fluence distribution of bright BATSE bursts assuming
that the GRB population follows the star formation rate.  Both groups concluded that their samples 
are consistent with the $E_{iso}-\nu_p$ relation discovered by Amati, but with a larger 
scatter than originally found (but cf., Nakar \& Piran, 2005).  

Here we are arguing that the scatter is to be expected.  
Clearly, if there are two effects that are creating two separate correlations,
a study of just one of them will find scatter.   As suggested above, both opening
angle effects (Frail et al. 2001) and viewing angle effects (EL04) are each creating 
their own correlation:  the opening angle effect would create scatter in $E_{iso}$ 
even for head-on viewers and even if there were no scatter in $E_\gamma$, and a viewing angle effect 
would create scatter in 
the observed $E_{iso}$ even if there were no scatter in $E_\gamma$ or in $E_\gamma/\theta^2$.
In addition, extreme outliers such as GRB 980425 can be attributed to a relatively large angle scattering
off material with a relatively modest Lorentz factor (Eichler \& Levinson 1999; Nakamura, 1998).
Assuming that this component is weak it can probably only be seen from very nearby GRBs.

In conclusion, a population of beamed sources with a standard energy 
output and a universal intrinsic spectrum, viewed along different sight lines, 
can explain the Amati and Ghirlanda relations.  The small scatter in the peak energy - 
collimation corrected energy correlation reported by GGL04 indicates that the
scatter in the $E_{iso}-\nu_p$ relation is predominantly due to a spread in the 
solid angle subtended by the gamma-ray emission region, and that the total energy 
output in gamma-rays of long GRBs is narrowly clustered.  The total energy released
as gamma rays should equal roughly the corrected energy $E_{\gamma,app}$ of the hardest 
source in the GGL04 sample, $E_{\gamma}\sim 2\times10^{51}$ ergs.  Consistency checks of the 
Amati and Ghirlanda relations on large samples should allow for the potentially 
large scatter produced by the opening angle effect and for biases in the beaming 
correction factor.  Larger samples of sources with measured redshifts, particularly 
dim soft bursts, are ultimately needed to confirm those relations.  The original argument of Frail
et al. (2001), that the scatter in $E_{iso}$ is greatly reduced when the beaming correction they make
is invoked may have naively appeared to have been confounded by the large spread in apparent 
$E_{iso}$ that remained.  Here, however, we have noted that the observable difference between the 
slopes of the Ghirlanda and Amati relations in fact provides an interesting confirmation to the 
afterglow theory invoked by Frail et al. (2001).  In particular, the noticable differences between 
the Amati and Ghirlanda slopes confirms the 
very weak dependence of opening angle on $E_{iso}$, which  may have otherwise gone without 
direct observational confirmation. 

This research was supported by the Israel-US Binational Science Foundation, 
an Israel Science Foundation Center of Excellence Award, and the Arnow Chair of Theoretical Physics.

\appendix
\section{Analytic derivation of the $E_{iso}-\nu_p$ relation}
Consider the emission from an annular jet centered around the $z$-axis, and having an opening 
angle $\theta_0$ and angular width $\Delta \theta$.  Denote by  {$ \hat \beta$} the 
directions of emitting fluid elements (assumed to have a fixed Lorentz factor $\Gamma$) 
and by  {$ \hat n$} the sight line direction, respectively,
and define  $\cos\theta_\beta=\hat{\beta}\cdot\hat{z}$, and $\cos\theta_n=\hat{n}\cdot\hat{z}$.
Then $\hat{n}\cdot\hat{\beta}=\cos\theta_\beta\cos\theta_n +\sin\theta_\beta\sin\theta_n\cos\phi$,
where $\phi$ is the azimuthal angle.  Assuming that each fluid
element emits isotropically in its rest frame a total energy
$E^\prime$, then its contribution to the overall energy emitted
into a solid angle $d\Omega_n$ around the sight line direction
$\hat{n}$, as measured in the Lab frame, is:
$dE(\hat{n},\hat\beta)=(E^\prime {\cal D}^3)d\Omega_n$, where
${\cal D}=1/\Gamma(1-{\bf \beta}\cdot \hat n)$ is the corresponding
Doppler factor. The fluence along the line of sight  is given by
the integral
\begin{equation}
E_{iso}=\int\frac{dE(\hat{n},\hat\beta)}{d\Omega_n} d\Omega_\beta\propto 
\int{\cal D}^3\sin\theta_\beta d\theta_\beta d\phi.
\label{Eiso-1}
\end{equation}
Define $\Delta=\theta_n-\theta_\beta$, and assume small angle approximation, viz.,
$\Delta\theta<\theta_0<<1$ and $\Delta<\theta_0$, the Doppler factor can be expressed
as:
\begin{equation}
{\cal D}(\Delta,\chi)=\frac{1}{\Gamma[1-\beta\cos\Delta+\beta\sin\theta_\beta\sin\theta_n(1-\cos\phi)]}
\simeq\frac{2}{\Gamma[\Gamma^{-2}+\Delta^2(1+\chi^2)]},
\end{equation}
where $\chi^2=2\sin\theta_\beta\sin\theta_n(1-\cos\phi)/\Delta^2$.  Define
$\eta^2=2\sin\theta_\beta\sin\theta_n/\Delta^2$, and 
\begin{equation}
F(\Delta)=\int_0^{\eta}\frac{d\chi}{(1+\chi^2)^3(1-\chi^2/2\eta^2)},
\end{equation}
equation (\ref{Eiso-1}) can be written, for angular separations $\Delta^2>>\Gamma^{-2}$, as
\begin{equation}
E_{iso}\propto\int_{\Delta_{min}}^{\Delta_{max}}\frac{8}{\Gamma^3\Delta^5}
\left(\frac{\sin(\theta_n-\Delta)}{\sin\theta_n}\right)^{1/2}F(\Delta)d\Delta,
\label{Eiso-2}
\end{equation}
with $\Delta_{min}=\theta_n-\theta_0$, $\Delta_{max}=\Delta_{min}+\Delta\theta$.  
In the limit considered here, viz., $\Delta<<\theta_0<\theta_n$, we have $\eta>>1$, so that
$F(\Delta)\simeq \int_0^\infty(1+\chi^2)^{-3}d\chi=3\pi/16$, and $\sin(\theta_n-
\Delta)/\sin\theta_n=1 -O(\Delta/\theta_n)$.  Equation (\ref{Eiso-2}) 
reduces to $E_{iso}\propto \Gamma^{-3}\Delta_{min}^{-4}$.  For a jet emitting an intrinsic 
single universal spectrum with a peak frequency $\nu^\star$, the observed spectral peak
will be located at
$\nu_{p}=\nu^\star{\cal D}(\Delta_{min},\chi=0)=2\nu^\star/(\Gamma\Delta_{min}^2)$
for sight lines outside the jet (EL04), whereby we obtain 
$E_{iso}\propto\Gamma^{-1}(\nu_{p}/\nu^\star)^2$.  Similar result can be readily derived for 
sight lines inside the hole.  Numerical integration of eq. (\ref{Eiso-1})
for a wide range of parameters and viewing angles is presented in EL04.

\end{document}